# A novel laser Doppler anemometer (LDA) for high-accuracy turbulence measurements

Mohd Rusdy Yaacob[1, ID]. Rasmus Korslund Schlander[2] . Preben Buchhave[3] . Clara Marika Velte[4, ID]

**Abstract**
High accuracy and dynamic range have been some of the most prominent challenges when it comes to fine-scale turbulence measurements. The current commercial LDA processors, which perform the signal processing of Doppler bursts directly using hardware components, are essentially black boxes and in particular are renown for suffering from practical limitations that reduce the measurement reliability and accuracy. A transparently functioning novel LDA, utilizing advanced technologies and up-to-date hardware and software has therefore been developed to enhance the measurement quality and the dynamic range. In addition, the self-developed software comes with a highly flexible functionality for the signal processing and data interpretation. The LDA setup and the combined forward/side scattering optical alignment (to minimize the effective measuring volume) are described first, followed by a description of the signal processing aspects. The round turbulent jet has been used as the test bed since it presents a wide range of degree of difficulty for the LDA processor (accuracy, dynamic range etc.) across the different radial distances and downstream development. The data are diagnosed for dynamic range in residence and interarrival times, and compared to a typical hardware driven processor. The radial profiles of measured mean streamwise velocity and variance agree well with previous studies of the round jet. The spatial turbulent kinetic energy spectra in the fully developed region perfectly match the expected (and in this region well established) -5/3 power law even for the largest measured distances from the centerline (where shear and turbulence intensity are significant).

**Graphical Abstract**

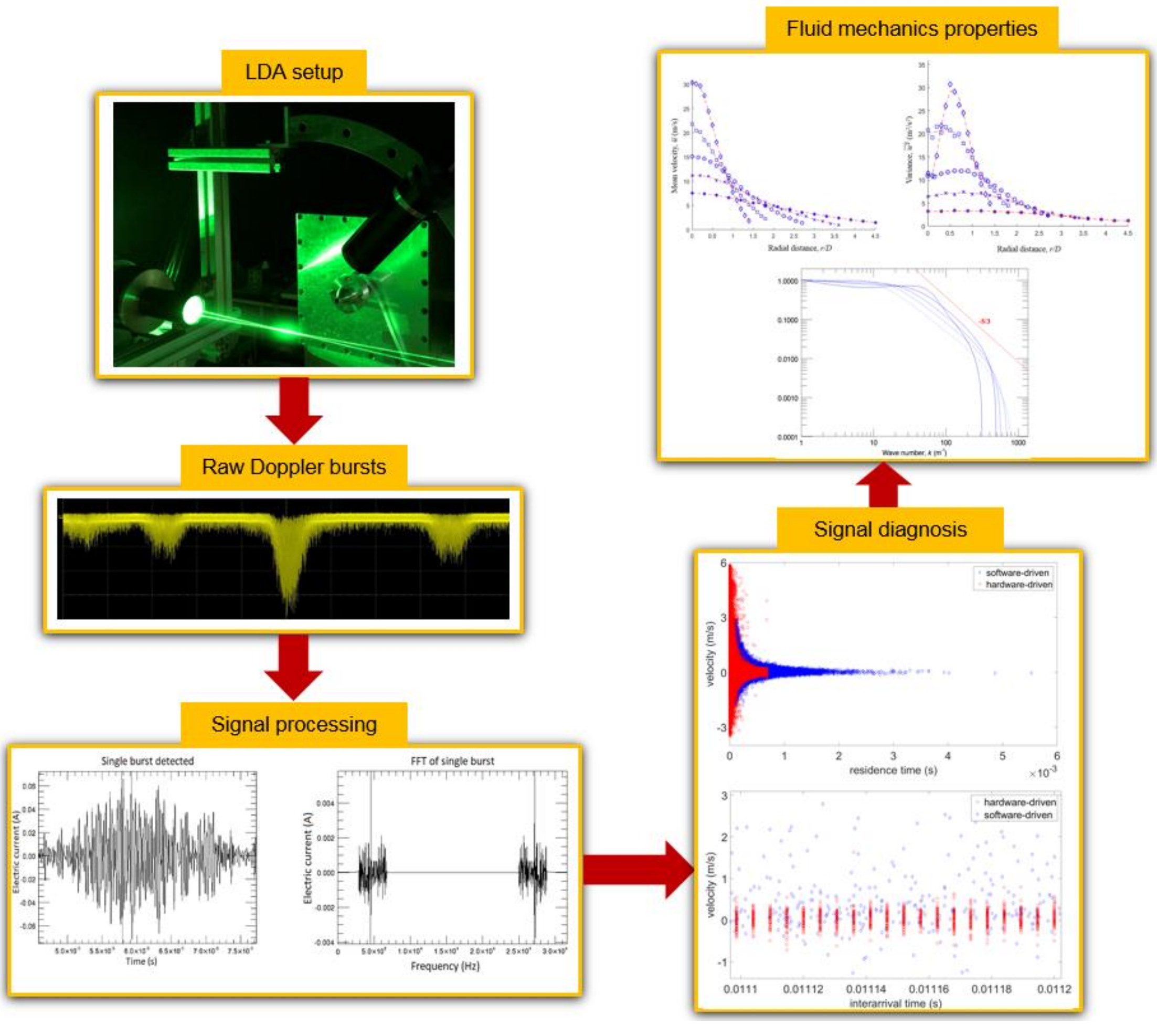

## 1 Introduction

Despite its omnipresence in various applications, e.g., combustion engines and weather forecast, turbulence has long been an area of classical physics which is yet to be completely defined and discovered (Davidson 2004; L'vov and Procaccia 2007). While the equilibrium aspects of stationary turbulence has been thoroughly investigated and understood (Batchelor 1953; Kolmogorov 1941c, 1962), the non-equilibrium counterparts are still largely underexplored. Vassilicos and his team (Goto and Vassilicos 2015; Mazellier and Vassilicos 2010; Valente and Vassilicos 2012; Vassilicos 2015) have raised serious questions to the established theory based on accumulated evidence from both experimental and simulation-based investigations.

One of the main reasons for the lack of our understanding of these non-equilibrium flows is that they are notoriously difficult to measure accurately since they are typically shear flows of high turbulence intensity and of great variations in dynamic range. If implemented and analyzed correctly, the laser Doppler anemometer (LDA in the following) is the most accurate existing instrument to measure these difficult flows, since it has inherently a wide dynamic range, unlike e.g., correlation based Particle Image Velocimetry, and it has the ability to, without ambiguity, distinguish between and discern the velocity components, unlike e.g., hotwires. Furthermore, a major challenge subsequently arises in measuring the kinetic energy spectrum, which is a powerful tool that can map the energy across all turbulence scales and provides valuable information about turbulence generation and its assumed cascade development more precisely (Frost 1977). Similar challenges arise when measuring the physical space counterparts, such as correlations (e.g., covariances) and structure functions (which are central to the nowadays debated Kolmogorov equilibrium description(s) of turbulence) (Kolmogorov 1941b, 1941a, 1941c, 1962).

LDA is one of the most preferred measurements techniques in these challenging turbulent flow measurement since it has been significantly recognized for its unique favorable properties in various experimental investigations (Barker 1973; Buchhave, George, and Lumley 1979; Peiponen, Myllylä, and Priezzhev 2009). It can truly distinguish the spatial velocity components of a flow from each other (Hussein, Capp, and George 1994) and therefore produce reliable data. However, existing commercial LDA systems have been restricted with some practical limitations, e.g., for turbulence measurements that require high dynamic range and signal-to-noise ratio (Buchhave, Velte, and George 2014; Velte, Buchhave, and George 2014). A more detailed discussion of the practical limitations can be found in (Velte, George, and Buchhave 2014).

It is therefore of great interest to develop a novel, well-functioning, LDA system that is able to measure turbulence more accurately. With a more well-functioning LDA processor, it is possible to credibly measure turbulence in experimentally challenging regions, e.g., non-equilibrium, high intensity and high shear regions, to test the debatable universal equilibrium theory of Kolmogorov (Batchelor 1953; Kolmogorov 1941b, 1941a, 1941c, 1962; Monin and Yaglom 1972).

## 2 Methodology

### 2.1 Flow generation facility

The axisymmetric turbulent round jet has been a popular research subject for turbulent flow investigations since many years (Abdel-Rahman, Chakroun, and AI-Fahed 1997). Moreover, in the fully developed region, it was proven to produce results that are in good agreement with the classical Kolmogorov theory of turbulence (Gibson 1962; Panchapakesan N. R. and Lumley 1993; Wänström, George, and Meyer 2012). At the same time, this flow presents a wide variation of degree of shear and turbulence intensities across the downstream and radial directions, which is why it has been chosen as the test bed for validation of turbulence measurement with our LDA system.

The axisymmetric turbulent round jet used for this measurement is a replica of the one used by Velte, George and Buchhave (2014). It is fitted with an outer nozzle at the end, having an exit diameter $D$ = 10 mm and contraction ratio of 3.2:1. Pressurized air is injected through the jet together with the glycerin particles (~1-5 µm) at regulated pressure values. The particles have been proven to be sufficiently small to faithfully track the flow to a sufficient degree, while also scattering light sufficiently well to be well detectable by the LDA system (Capp 1983). The jet is mounted on a two-axis traversing system which is driven by, for each axis, a hybrid two-phase stepper motor and computerized by a motion control software, *RemoteWin* for maneuvering the jet along streamwise ($x$) and radial ($r$) directions. With this, the jet centreline can be easily traversed to different coordinates within the flow for turbulence measurement at high spatial resolution. The choice of traversing the jet instead of the LDA optics is based on the sensitivity in alignment of the LDA optics, since this is a combined forward/side scattering system (as will be described below).

### 2.2 Laser Doppler anemometer

The LDA is operated in the burst-mode (Roberts, Downie, and Gaster 1980) and consists of a continuous wave laser beam with wavelength, $\lambda$ = 532 nm split into two coherent beams. The two beams are directed through a dual Bragg cell (BC in the following) in order to distinguish the moving direction of the particles along the measured component axis (Buchhave 1984; Diop, Piponniau, and Dupont 2019). This issue was remarked to be critical, in particular within the flow region where fluctuations and turbulence intensities are high (Yaacob et al. 2018), since the velocity variations are more likely to exist in both directions. The frequency of one of the beams is shifted by 40 MHz, which value is known from the BC module used, while the frequency of the other beam is shifted by 37 MHz. The two beams consequently experience 3 MHz of effective frequency shift, $f_s$ which creates movement of the interference fringes in the direction opposite to the main flow direction. This value is sufficient for acquiring the maximum Doppler frequency or velocity from our measurement (Peiponen, Myllylä, and Priezzhev 2009).

The parallel beams are then passed through a converging lens and focused to intersect at a focal point of 200 mm based on the dual-beam principle (Bartlett and She 1976; Grant and Orloff 1973; Sommerfeld and Tropea 1999). The volume where the frequency shifted beams intersect, commonly referred to as the measurement volume (MV in the following), is where the local velocity of the flow is measured (Albrecht et al. 2003).

The working block diagram for acquiring the data from LDA is depicted in Fig. 1. The photodetector receives the light scattered by each seeding particle and converts the photons into photocurrent. The detector is also coupled to a photomultiplier (Hamamatsu H10425) that amplifies the photocurrent internally before passing through the load resistor in the filter circuit. A first-order low pass filter (load resistor, $R$=270 Ω) is connected between the photomultiplier which has capacitance value of 22 pF, and the amplifier, giving a resulting cut-off frequency of around 26 MHz. The analog frequency modulated signal is then delivered to and visualized through a high end oscilloscope. An A/D converter is also embedded in the scope yielding a 13 bits resolution. In order to utilize this resolution, the amplitude on the oscilloscope must be set such that most of the bursts can be seen in their full size. Some of the largest burst are allowed to be clipped in order to get most of the bursts digitized with the full resolution, which is critical in computing the energy spectra. The clipping was shown not to cause any biasing of the Doppler frequency.

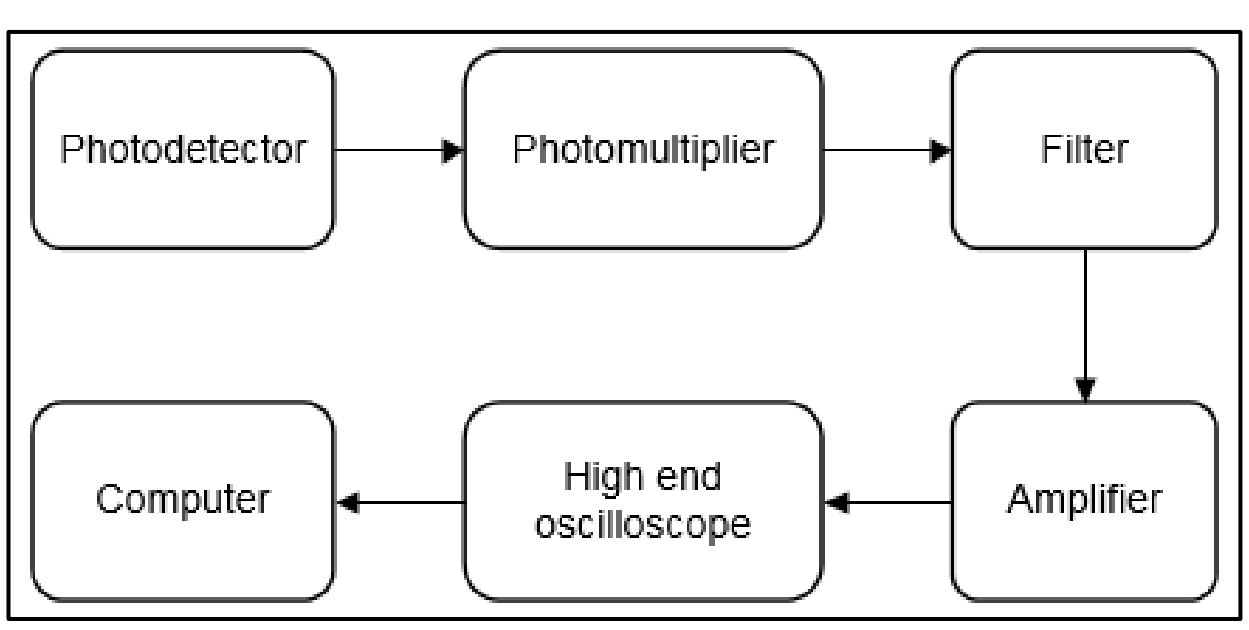

**Fig. 1** Block diagram for data acquisition of the LDA system

The experimental setup (see Fig. 2) is enclosed in a large tent of dimensions 3 x 5.8 x 3.1 m$^3$, made out of black canvas to minimize light pollution and to create an undisturbed environment for the jet to freely develop. With the jet positioned at the back of the enclosure, the jet flow generated in the facility should be expected to correspond well to a free jet up until $x/D$ = 70, which is sufficient for the purpose of our current experiments and future investigations (Hussein, Capp, and George 1994; Wänström, George, and Meyer 2012). The LDA system is operated in a forward-scattering detection mode by mounting the detector at 45° from the MV in order to minimize the light extinction from the Mie scattering and improve the signal-to-noise ratio (Fischer 2017). Such configuration will also result in a smaller and nearly spherical MV (since the optical cross-section of the detector dictates the effective MV) to obtain unbiased measurement especially in a highly turbulent flow and achieve the highest possible spatial resolution (Buchhave and Velte 2017b). A schematic side view of the setup is also shown in Fig. 3.

### 2.3 Optical alignment

Operating the system in a combined a forward/side scattering mode demanded thorough and rigid adjustment on the optical parts in order to assure good quality of the Doppler signals from the measurement. It is difficult to see by naked eye whether the two beams do in fact overlap to produce an MV. Therefore, a microscope objective is used to produce magnified beam spots on the wall of the tent (see Fig. 4). The beams are aligned accordingly to make them overlap to a good approximation. Precedent to this alignment, the MV was first assured to be at the distance equal to the focal length of the focusing lens, i.e., 200 mm. Apart from that, the working distance between the detector and the MV should also be determined by the focal length of the focusing lens (Tropea, Yarin, and Foss 2007), i.e., 200 mm, or at 1:1 ratio to the focal length.

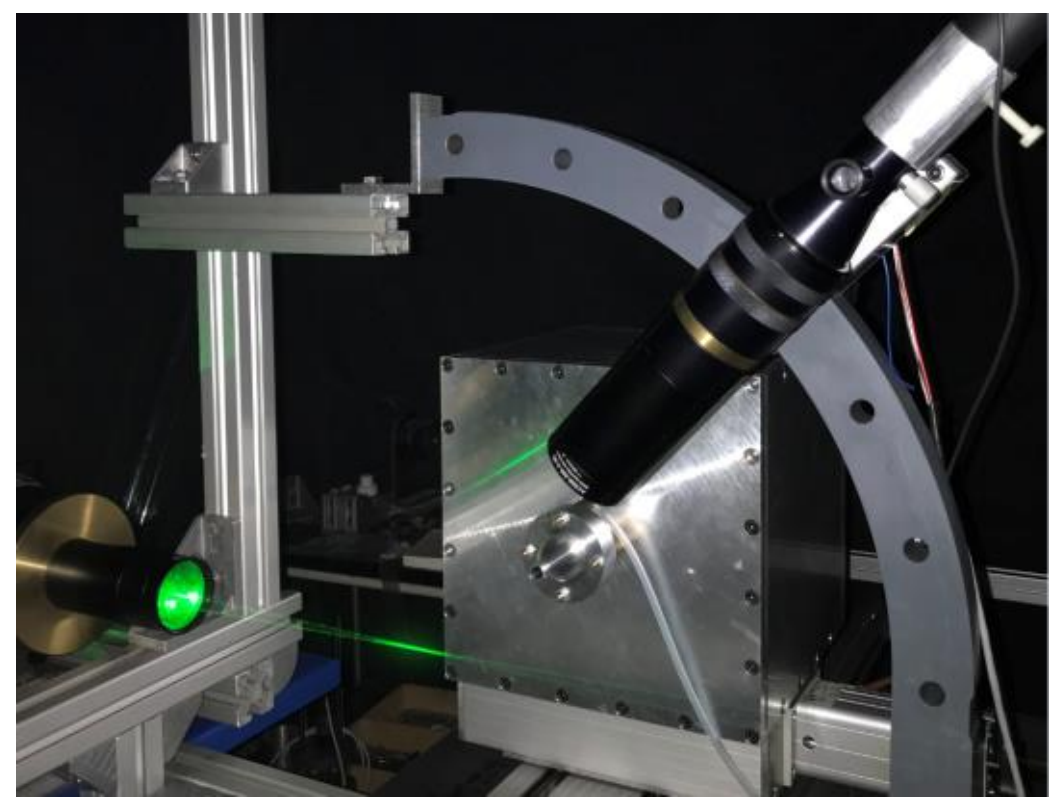

**Fig. 2** LDA system in a large tent (3 x 5.8 x 3.1 m$^3$)

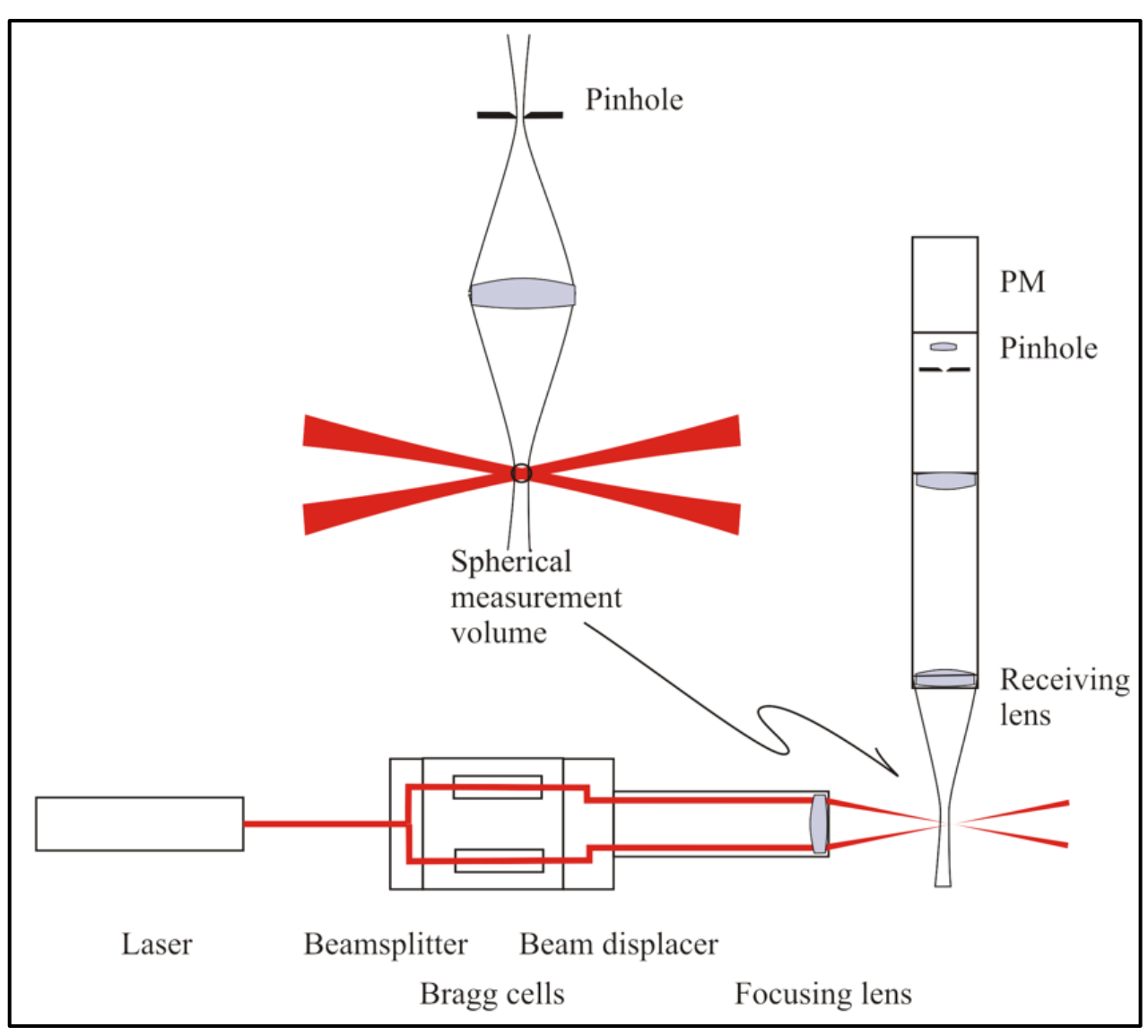

**Fig. 3** Schematic drawing (side view) of the LDA setup, displaying a 90° (side scattering) configuration. However, the detector is mounted at 45° from the MV during the measurements.

In addition, the photodetector must be well aligned with the MV. By looking through the photodetector, there is a black pinhole located in the middle of a circular area. The photodetector, which is mounted to a holder, needs to be adjusted so that the pinhole resides at the center of the beam intersection as in Fig. 5. However, a more reliable test of signal quality is to simply observe the burst signal from an analog scope which is temporarily channeled from the output of the amplifier. The pinhole's position at which the highest burst S/N ratio is observed on the scope should be chosen for the measurement. The amplifier is also set to an optimum value in a way that the high value of current setting should not cut the burst off too much.

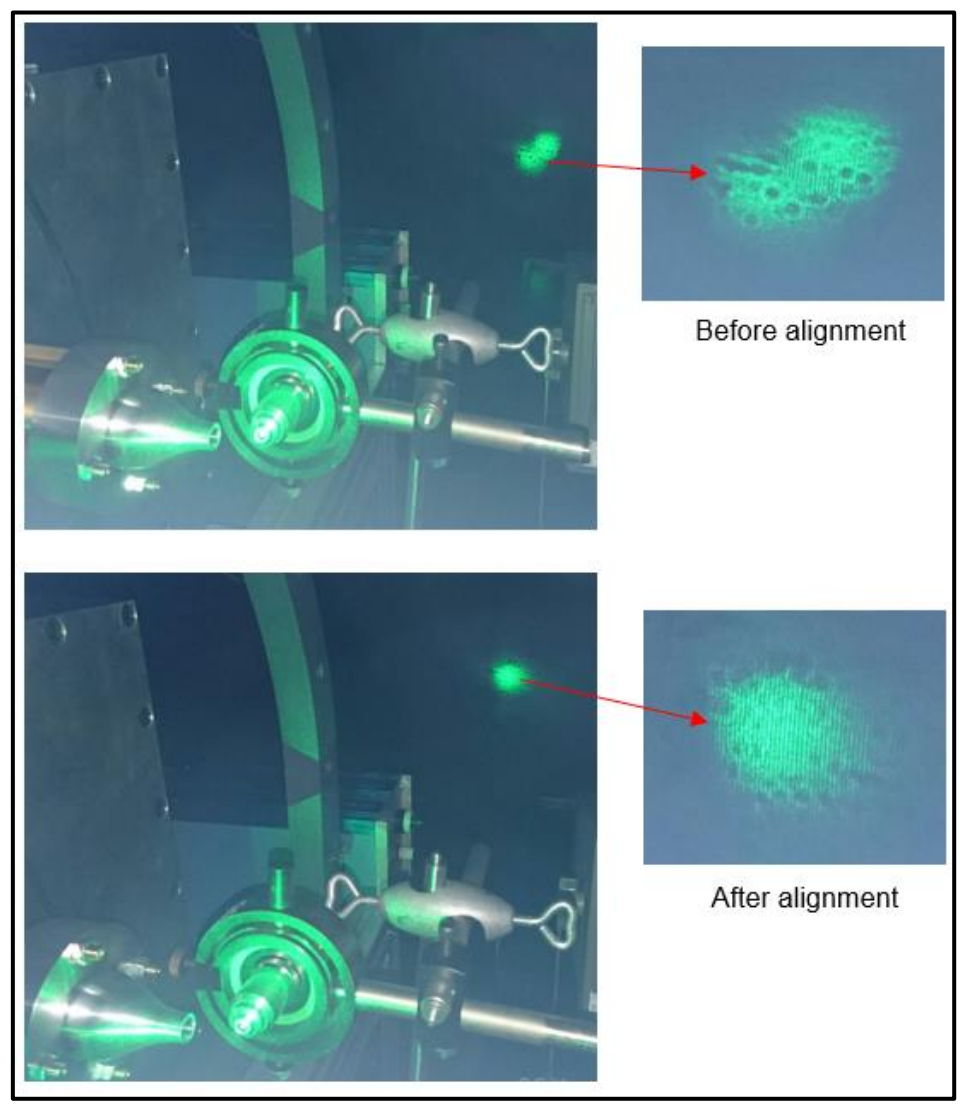

**Fig. 4** Alignment of the beams' overlapping

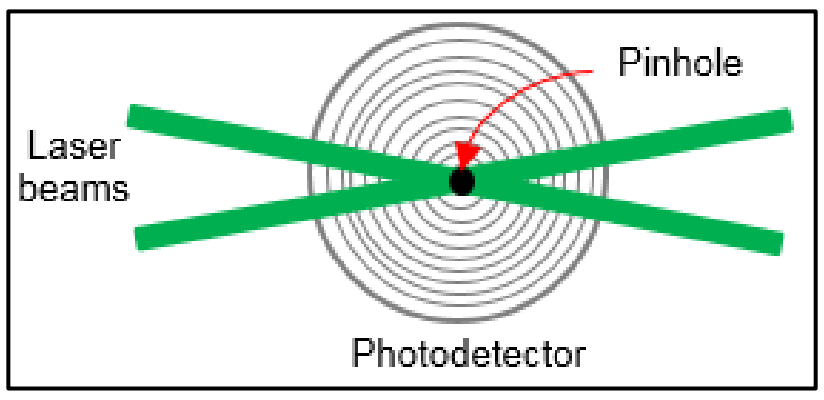

**Fig. 5** Pinhole alignment with MV through the photodetector

### 2.4 LDA measurement

A series of measurements spanning from $x/D$=5 up to $x/D$=30 downstream from the jet exit were carried out with the LDA system. For each downstream position, measurements were acquired at several points in the radial direction across the shear region, as depicted in Fig.6.

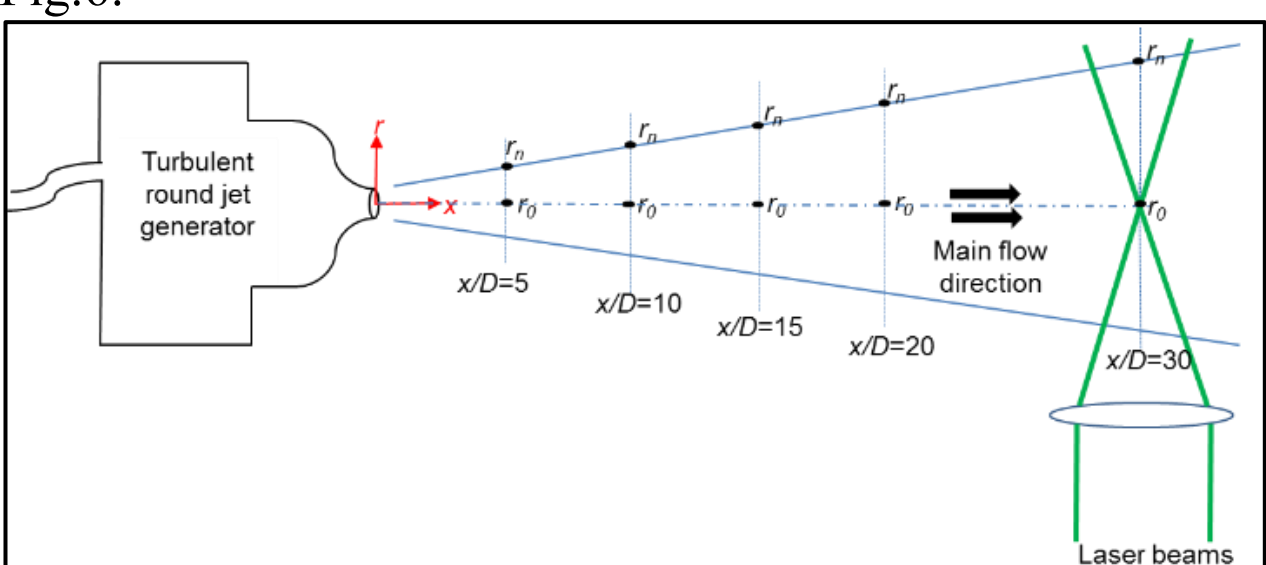

**Fig. 6** Schematic of the measurement points (MP), which are listed in detail in Appendix I.

The jet input pressure was set to give an exit velocity, $U_0$ of 32 m/s ($Re$=25000). Seeding particles were fed into the flow at a pressure of 1.2 bar which resulted in an optimum number of Doppler bursts (spatial seeding density resolving the relevant scales without significant burst-overlap) as seen from the scope (see Fig. 7). The seeding was allowed to distribute uniformly throughout the ambient air to improve the homogeneity of the seeding. The laser was adjusted to nearly reach its maximum intensity (1.29 W), along with 70 µA amplifying current on the photomultiplier.

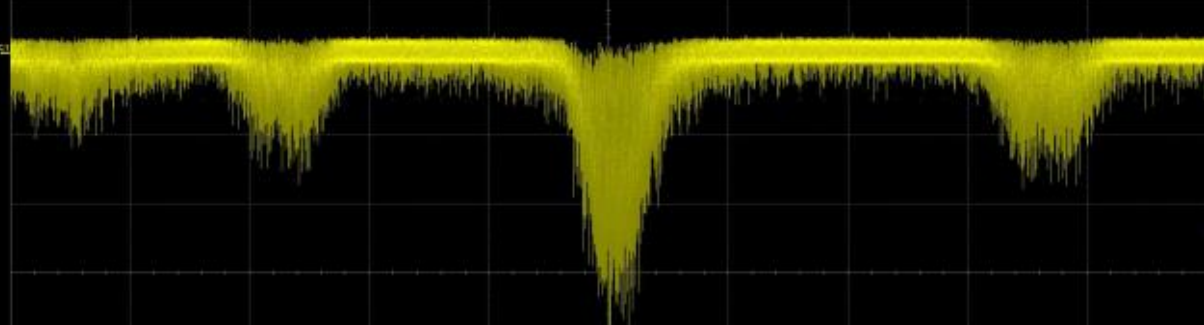

**Fig. 7** Doppler bursts acquired from the scope

All the abovementioned parameters were consistently used throughout the whole measurement except for the sampling rate of the raw burst-signal and the record length, which were optimized to accommodate to the flow variations. The sampling rate of the raw burst signal was chosen to fulfill the Nyquist sampling rate in relation to the Doppler frequency. The maximum frequency was determined from the built-in spectrum analyzer of the scope as in Fig. 8, prior to data recording at every measurement point. A sampling rate of either 12.5 MHz or 25 MHz was chosen for all measurement points, accommodating to the variations in flow conditions at each measurement point. The resulting record lengths were 2 s and 1 s, respectively, which are long enough to obtain sufficient statistics from the flow. A total of 400 records were taken at each measurement point.

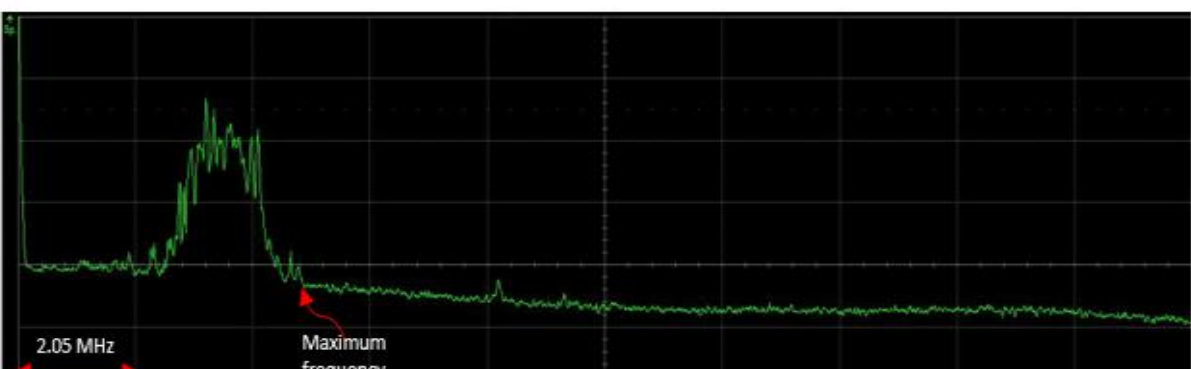

**Fig. 8** Range of Doppler frequency at one particular measurement point

### 2.5 Signal processing

The output measured at each measurement point is an equidistantly sampled digital record of electrical current and the acquired signal is in the form of an array of discrete points. Beside all the measured Doppler bursts, the signal also contains various sources of noise, e.g., quantum noise from the photomultiplier, thermal noise from the detector's circuit or optical noise from the surroundings (Buchhave, George, and Lumley 1979). This noise needs to be minimized to clearly reveal (to the processor) the desired burst. By transforming the signal to frequency space, most of the frequencies are clustered around a range (see Fig. 9) with frequencies corresponding to the measured velocities. Anything outside this range is considered as unwanted noise and therefore not to be processed. By using the calibration factor, $d$ (from $d = u/f_D$, where $u$ is the local instantaneous flow velocity and $f_D$ is the Doppler frequency), and since the approximate velocity is known, it is possible to remove potentially unphysical velocities/Doppler frequencies. In this case, the frequencies in the center correspond to velocities between -40 to 40 m/s, and anything outside of this range is deemed unrealistic.

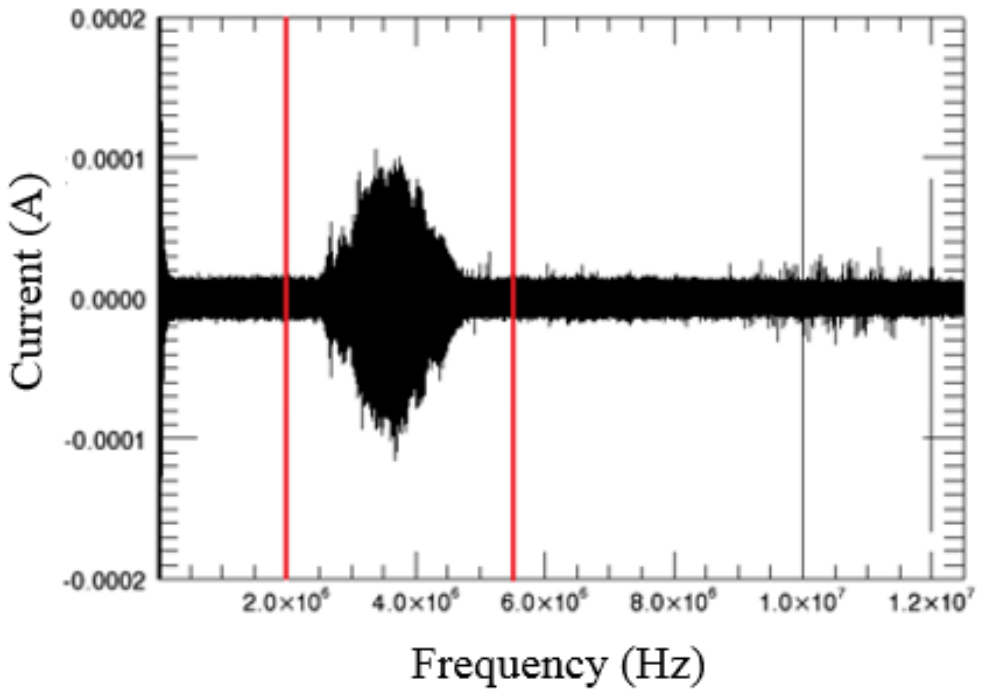

**Fig. 9** Simple band pass filter (band within the red lines) to isolate Doppler bursts

In order to detect the individual Doppler bursts, it is necessary to find their envelope. This is done by using the Hilbert transform (Schneider et al. 2007), which shifts the function by $\frac{\pi}{2}$. A sum signal is created, which is the original signal squared added to the Hilbert transformed signal squared. This yields an envelope of the signal, which is shown in Fig. 10. However, the signal is still too noisy to detect the individual bursts properly. Therefore, the envelope signal is filtered by a running convolution that is adjusted to remove noise at frequencies above the minimum burst length. The frequency is chosen to avoid bias, which could otherwise reduce the number of high velocity bursts. The result of the convolution is shown in Fig. 11, which also illustrates two horizontal lines known as trigger lines that are used to detect every burst having an amplitude higher than the upper trigger line. The burst is recorded until the amplitude gets below the lower trigger line. This method, commonly known as Schmidt triggering, is used to reduce the probability for detection of a noise spike. After detection, each burst is analyzed using the fast Fourier transform (FFT) in order to find the Doppler frequency. The challenge is now to extract a single frequency in finding the corresponding particle velocity**.** A single burst is shown in Fig. 12. Even though the Doppler burst is fairly visible, there are still many frequencies present throughout its extent.

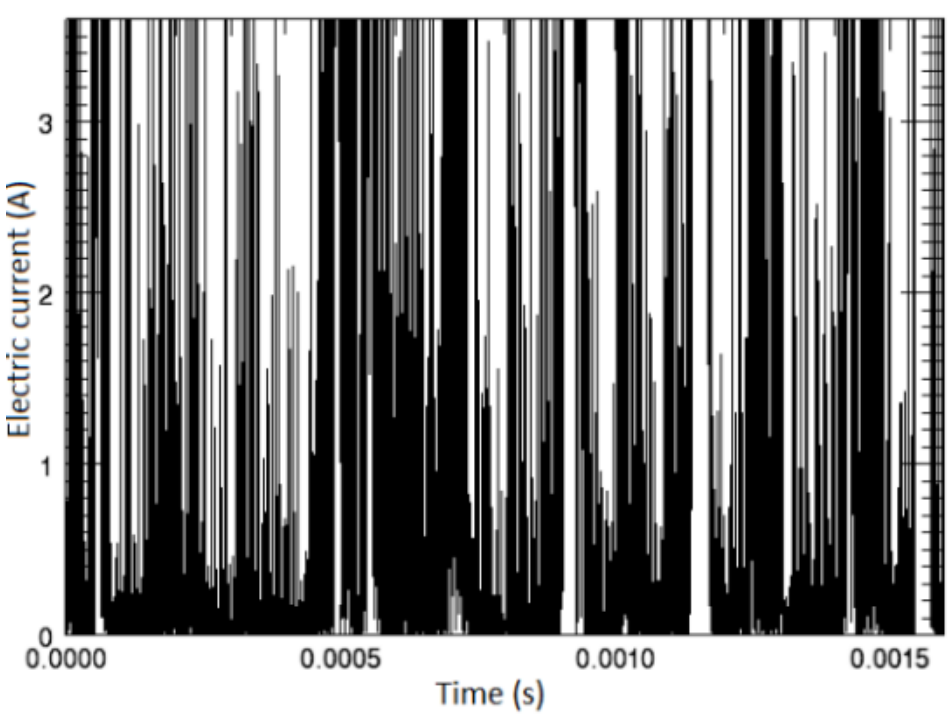

**Fig. 10** The sum signal

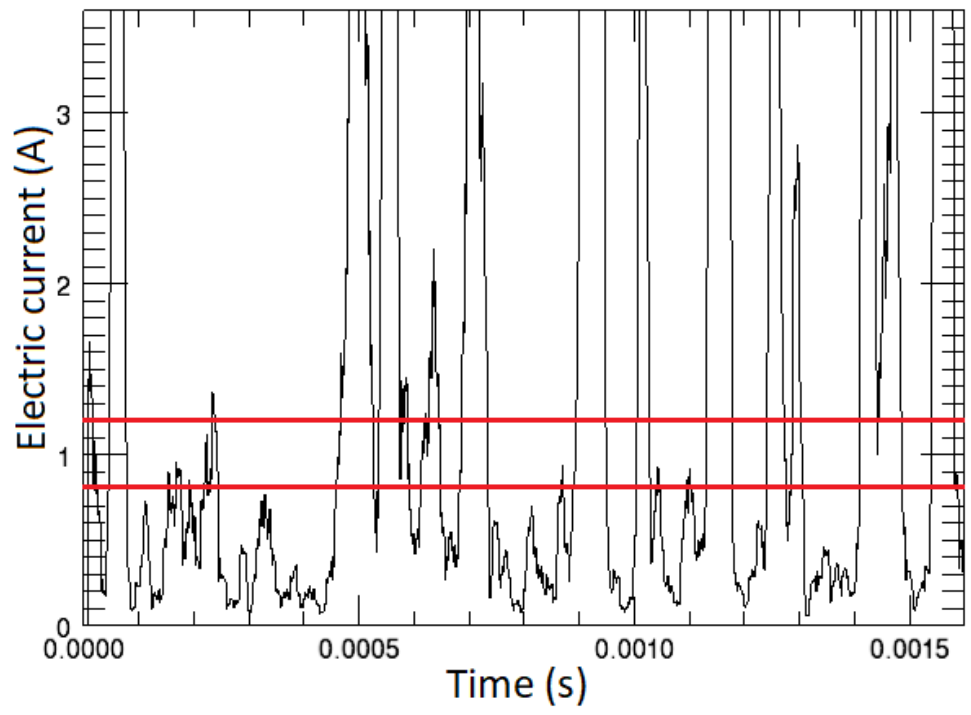

**Fig. 11** Convoluted sum signal now used to detect Doppler bursts using a Schmidt trigger system (illustrated by the two horizontal red lines)

In order to find the corresponding velocity from each Doppler burst, the FFT is applied to each burst as shown in Fig. 13. Only one of the obtained frequencies should correspond to the velocity for the particle moving through the MV. The situation is further complicated by the fact that, in addition to frequencies due to the noise, the Doppler frequency itself can change across the burst as the seeding particles may have a varying velocity when transiting through the finite sized MV. This is natural, since the spatial velocity gradients in turbulence are expected to become larger for smaller flow scales. A Gaussian function is therefore fitted across the Doppler peak and its maximum value is used as the frequency of the Doppler shift (see Fig. 14). This method is effective since the Doppler burst has the shape of a Gaussian function, and transforming such a function to Fourier space, will result in a Gaussian function as well.

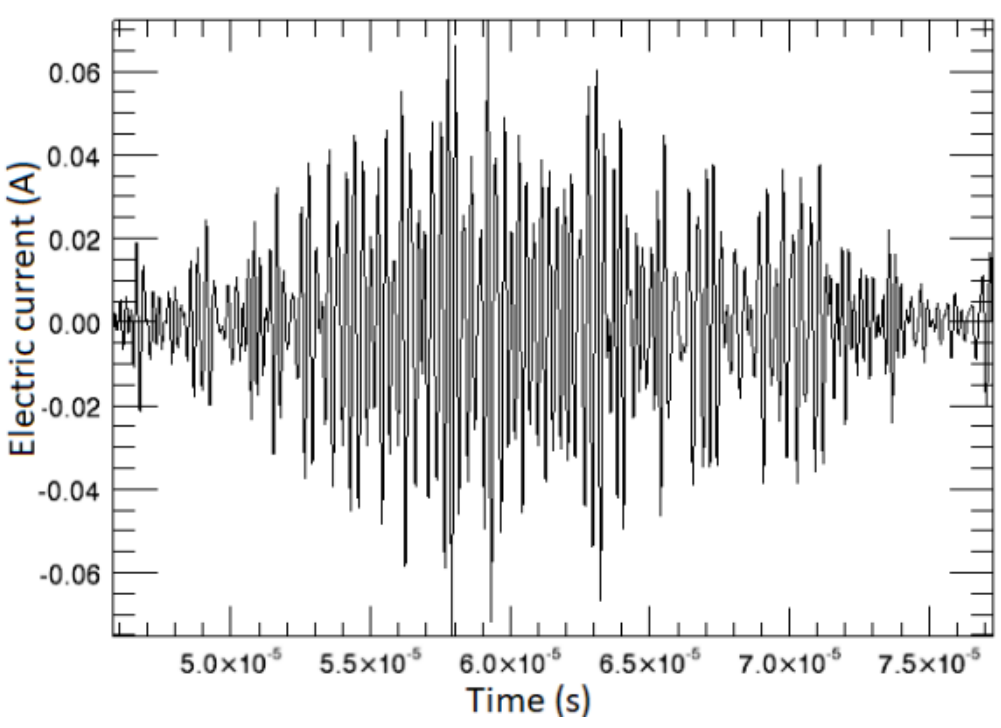

**Fig. 12** An example of a filtered Doppler burst with several frequencies contained

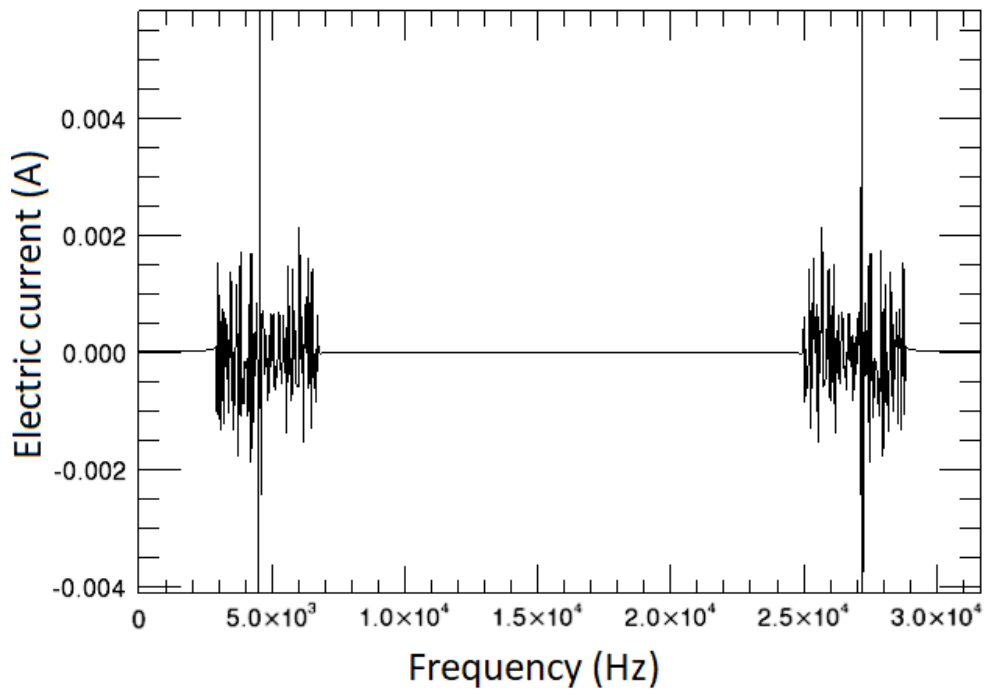

**Fig. 13** FFT of a single (band pass filtered) burst

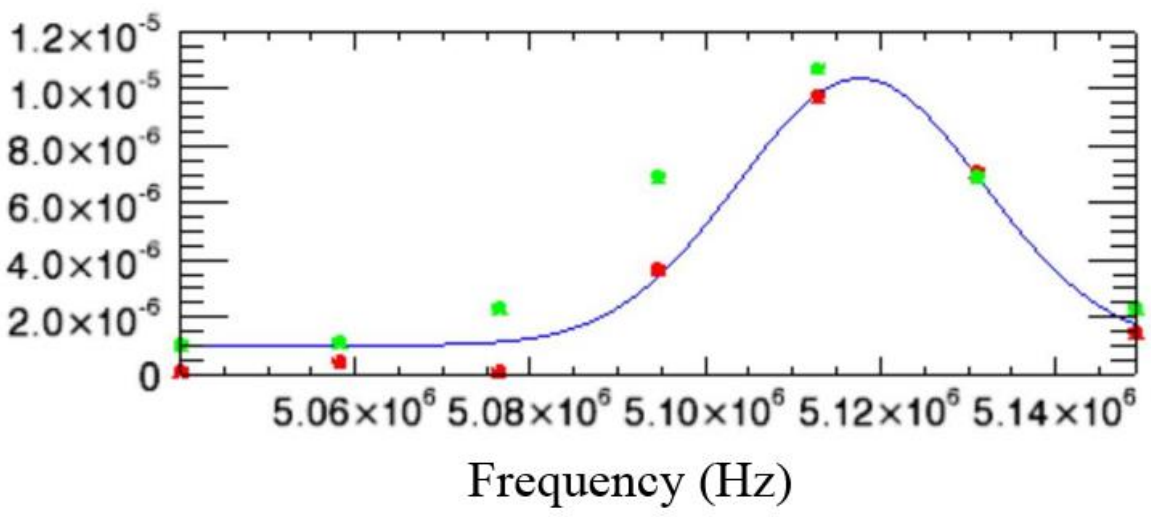

**Fig. 14** Gaussian fit across the Doppler peak. *Red square* ■ : measured points. *Green square* ■ : suggested function for the fit. *Blue curve*: best fit to the measured points.

### 2.6 Diagnosis of processed signal

Since our software driven LDA processor was initially developed to overcome the practical limitations of the hardware based ones (Velte, George, and Buchhave 2014), the processed signals from both types of processors will be simultaneously diagnosed. For this purpose, the data acquired from the commercial LDA system by Velte, George, and Buchhave (2014) are taken into consideration.

A central problem with the commercial hardware driven systems that the authors have experienced with is that both the long and short residence times are not well-represented in the data set due to the hardware limitations in the burst sampling. An accurate representation of these residence times is necessary for measuring unbiased statistics, in particular for high intensity and high shear flows, as has been shown from first principles (Velte, George, and Buchhave 2014). This misrepresentation of the residence times should not be problematic using this software driven system, which is highly flexible in this regard.

This can be illustrated with a simple diagnostics tool for LDA data quality, namely by scatter plotting the instantaneous velocities against their respective residence/transit times. For large positive convection velocities and relatively low turbulence intensities, the scatter plot should take on the familiar 'banana' shape, where high velocities are represented by small residence times and low velocities by long residence times (Capp 1983). For average convection velocities close to zero, the data should be distributed evenly around the residence time axis and reach high values. The traditional processors are clearly limited in this respect, as was seen in (Velte, George, and Buchhave 2014).

This is illustrated in Fig. 15, where the novel software driven processor is compared to a commercial counterpart in a turbulent round jet at 30 jet exit diameters downstream and two different radial distances; $r$=26 mm and 52 mm. The slight differences in global scatter distribution appear due to different validation of measured data between the processors as well as slightly different flows as similar, but different, jet generators have been used. This is, however, not critical to the current comparison.

Much longer residence times can be captured at the lowest (near zero) velocities by our novel processor compared to that of the commercial one, which maximum value is abruptly limited at the outermost off-axis positions (52 mm). The length of the longest measurable bursts as defined in the commercial hardware has effectively limited the maximum burst length and consequently the quality of low mean velocity (high intensity) result. One may optimize the hardware driven LDA measurements with improved settings, but the inherent problem with long residence time clipping is in principle better countered using the flexible software driven processor.

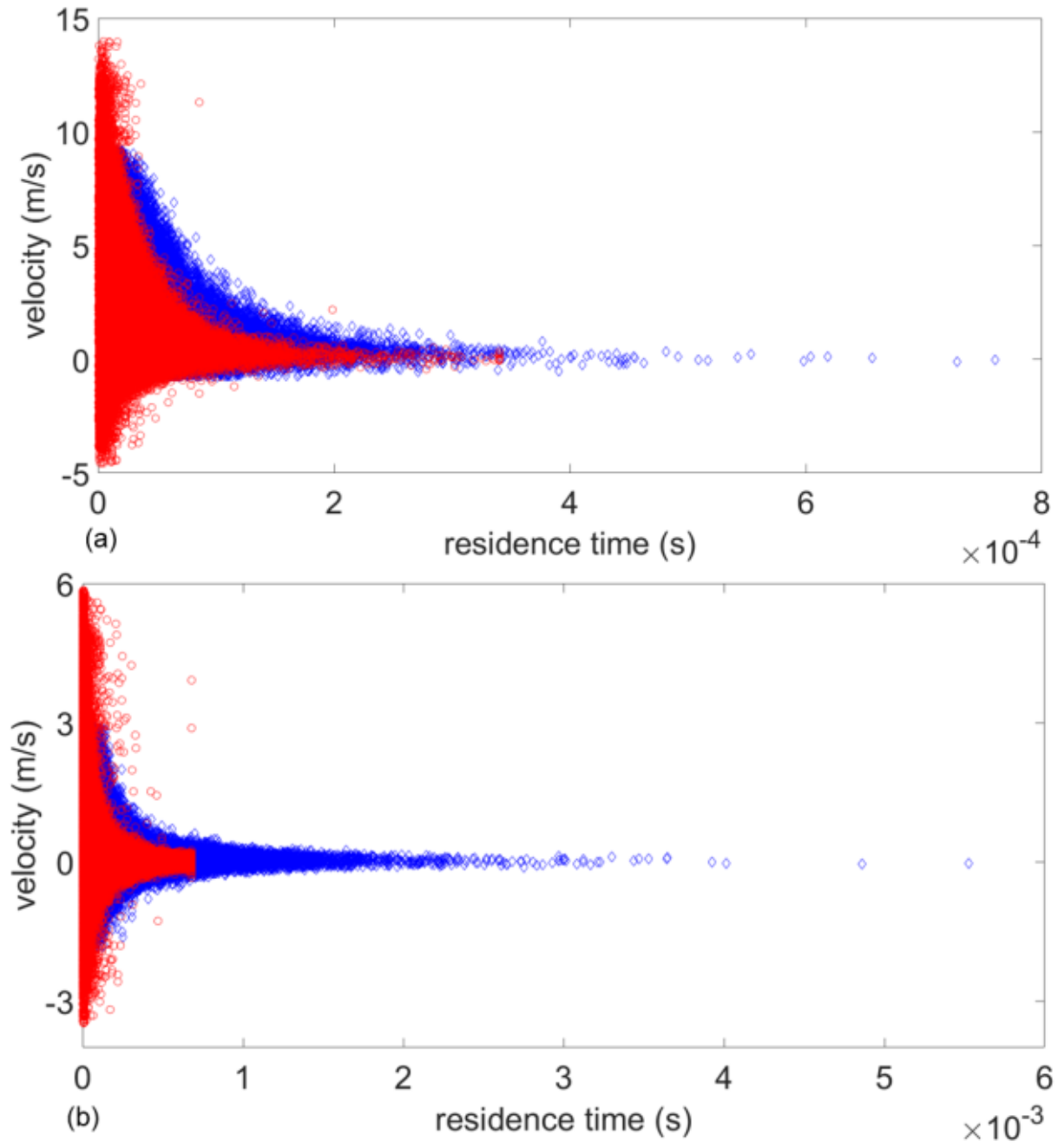

**Fig. 15** Velocity vs. residence time scatter plots for **a** $r$ = 26 and 27 mm, **b** $r$ = 52 and 51 mm. *Red circle* ○ : hardware driven LDA processor, *Blue diamond* ◇: software driven LDA processor

Another critical effect due to digitization could be found in the interarrival times (Buchhave, Velte, and George 2014; Velte, Buchhave, and George 2014). A closer examination showed that the processor had a finite data transfer time for each measured burst, which effectively limited the minimum attainable time between measurements. This can be illustrated, e.g., by a zoomed-in scatter plot of the individual velocities against the difference in arrival time between neighboring bursts (see Fig. 16). The vertical lines and the separations between them show the digitization and dead time effect, respectively. In power spectra computed using the residence time weighted discrete Fourier transform (DFT), this dead time was shown to produce oscillations in the high frequency end of the spectrum (Velte, George, and Buchhave 2014), originating from the Fourier transform of the dead time window. These regular interarrival time intervals appear to originate from data transfer times, during which new measurements cannot be acquired.

This dead time introduced has consequently been reported to have been removed in newer generations of the LDA processors for some of the commercial producers (private communication). However, the dead time effect of the finite measuring volume (no more than one particle allowed in the MV at any time when operating in burst-mode), which is a conceptual limitation of the optics, apparently must remain also in the current processor. The reduced effective MV size from operating the LDA in the combined forward/side scattering configuration does however aid in reducing this problem.

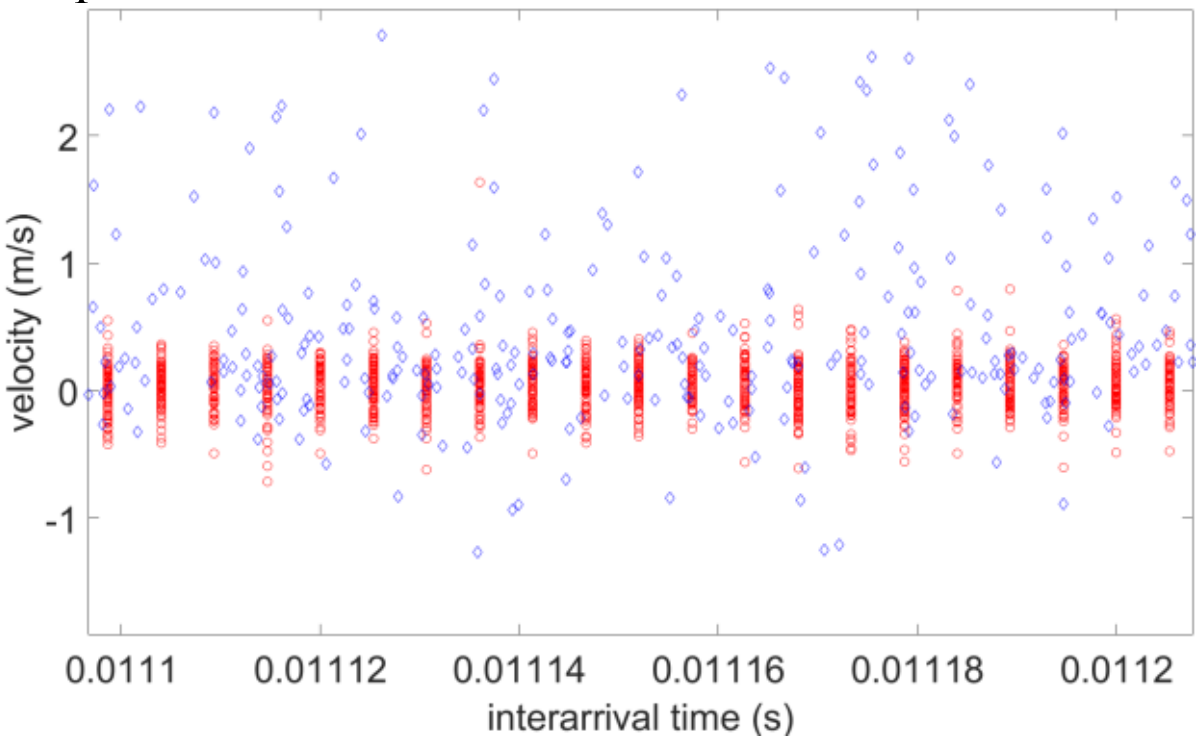

**Fig. 16** A zoomed-in of the velocity and the difference in arrival time between neighboring points. *Red circle* ○ : hardware driven LDA processor at $r$ = 52 mm, *Blue diamond* ◇ : software driven LDA processor at $r$ = 51 mm

### 2.7 Analysis

The output of all previous steps results in a temporal record of randomly sampled velocities, $u_i(t)$, since the seeding particles enter the MV randomly (the random arrivals of the particles in the MV effectively dictate the sampling) (Lading, Wigley, and Buchhave 1994). With optimal seeding density (data rate), one can maximize the highest frequency (smallest scale) resolvable, while keeping the amount of overlapping bursts (more than one particle in the MV simultaneously) to a minimum. This is important in getting a higher dynamic range of the power spectrum (Buchhave and Velte 2015). The random sampling prevents the use of the FFT, which requires equidistant samples. Instead, the discrete Fourier transform (DFT) must be used, which computes the transform at the actual (random) sampling times. This manner of computing the Fourier transform is usually slower than the FFT. However, we have developed a fast array processing algorithm for the DFT which makes it comparable in computational speed to the FFT (Buchhave and Velte 2015; Velte, George, and Buchhave 2014). The velocity power spectrum can now be computed using Equation (1):

$$S_i(f) = \frac{1}{T}\hat{u}_i(f)\hat{u}_i(f)^* \tag{1}$$

where $T$ is the length of the time record and $\hat{u}_i(f)$ is the Fourier transform of $u_i(t)$. For randomly sampled data, computing the Fourier transform of the velocities (or any statistics for that matter) requires special care and, as has been shown from first principles, should be carried out using residence time weighting to obtain unbiased statistics (Buchhave 1979; Buchhave, George, and Lumley 1979; Buchhave and Velte 2015; Buchhave, Velte, and George 2014; Velte, Buchhave, and George 2014; Velte, George, and Buchhave 2014). The spectra display the kinetic energy distributed across the measured bandwidth of frequencies.

To avoid scrambling of energy due to the fluctuating convection velocity and similar effects (Buchhave 1979; Lumley 1965), the energy spectra presented in the Results part are plotted in the wavenumber, $k$, domain:

$$S_i(k) = \frac{1}{L}\hat{u}_i(k)\hat{u}_i(k)^* \tag{2}$$

where $L$ is the length of the spatial record and $\hat{u}_i(k)$ is the Fourier transform of $u_i(s)$. The spatial record has been computed based on the convection record method (Buchhave and Velte 2017a) which does in an exact manner what Taylor's hypothesis only approximates (Buchhave and Velte 2017b). Thereby, the spectral scrambling and other adverse effects of the time spectra are effectively avoided.

## 3 Results and Discussions

Previous studies have shown that the mean velocity profile of a typical fully developed turbulent round jet flow should follow the well-known Gaussian distribution (Ball, Fellouah, and Pollard 2012; Mossa and Serio 2016). This behavior is also observed in the results obtained from our measurement (see Fig. 17) which covered only one half of the jet (up to a maximum of ~45 mm in the radial direction), since we assumed symmetry. Note that, the radial distance is normalized by the exit diameter of the jet. For each profile, the mean velocity is the highest at the jet centerline and approaching zero at large distances away from the centerline, as expected. The shape is also tapering off in the upstream direction. Meanwhile, as also expected, the streamwise velocity variance profiles (see Fig. 18) indicate the positions of maximum shear at each highest value, from which they also spread and taper with the downstream development.

As a quantitative test of the accuracy of the LDA velocity measurements, it is illustrative to test momentum conservation of the round turbulent jet. Momentum is conserved for a jet in an infinite environment, which is well approximated here for a jet exit diameter $D$=10 mm and an enclosure of 3 x 5.8 $m^2$. The velocity moment profiles can be tested using the momentum integral approximated to second order in (Hussein, Capp, and George 1994), which is valid in the fully developed jet region. We find that the ratio between the momentum flux per unit mass at $x/D$=30 to that at the jet exit is $M/M_0$ =0.99. More details on the calculations can be found in Appendix II. This shows that the profiles of the velocity moments obtained at $x/D$=30 from Fig. 17 and 18 satisfy momentum conservation, as expected. These convincing results also support that our LDA system is well suited for this kind of highly challenging measurement especially in the outer jet where velocity fluctuations are large, which demands a high dynamic range to accurately measure the small velocity changes (Jensen 2004).

The shape of the kinetic energy spectrum was also previously investigated in the fully developed (equilibrium) region (Fellouah, Ball, and Pollard 2009;

Gibson 1962), which for the current nozzle corresponds to approximately *x/D*=30 or beyond, where turbulence has become fully developed. For a clear comparison, each spectrum has been normalized with their respective mean square velocity value in Fig. 19. The results show good agreement with the expected -5/3 power law for an (assumed) inertial range (Batchelor 1953; Gibson 1962). Previous experimental investigations have been done by Velte, Buchhave and Hodzic (2017) using stereoscopic Particle Image Velocimetry, which results are clearly consistent with the present ones. The range within which each spectrum follows the -5/3 slope is also significant from our results, even for large radial distances from the centerline, strongly supporting that our novel LDA system is highly reliable even for high intensity and shear turbulence measurement.

Spectra at the centerline for every downstream position are also shown in Fig. 20, which demonstrates significant deviations from the power law as the position of measurement moved away from the equilibrium region.

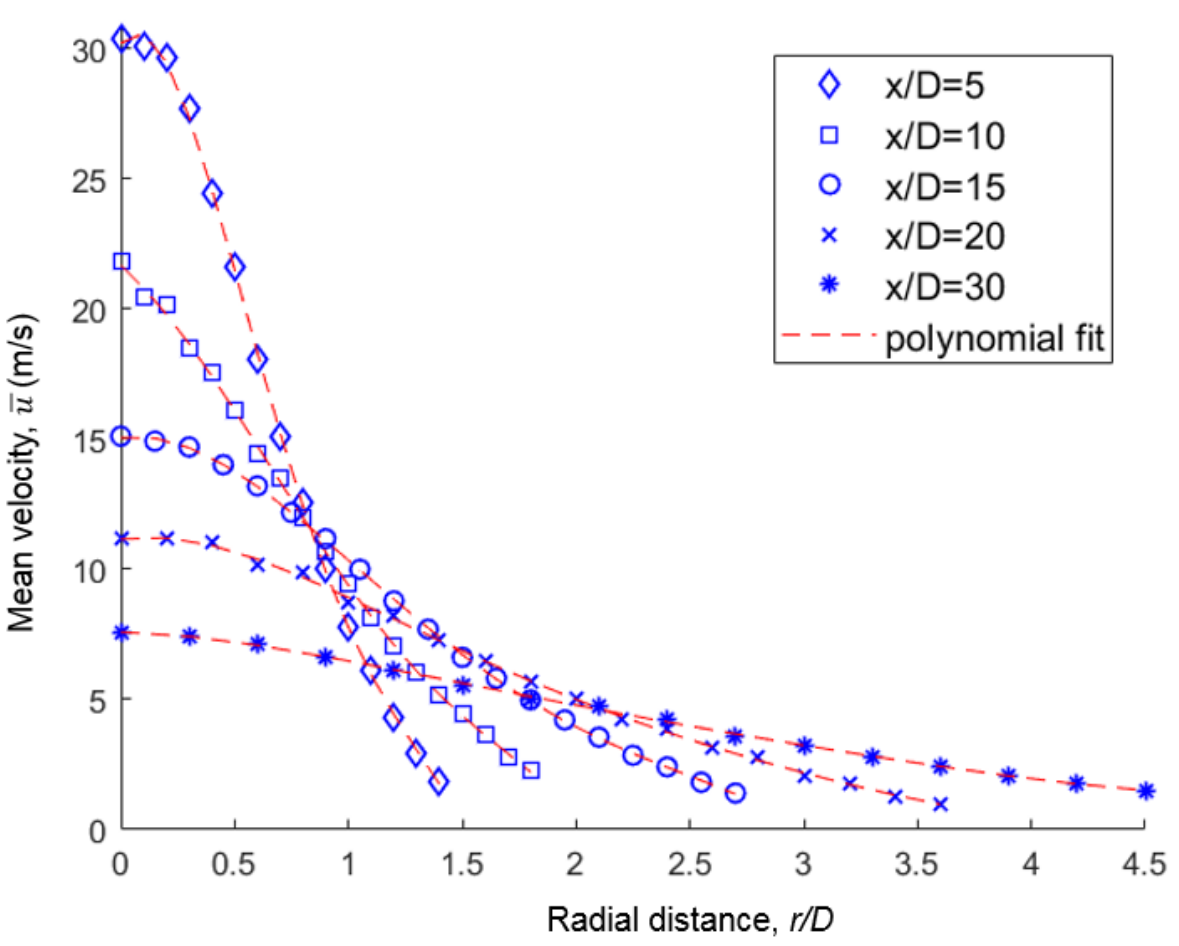

**Fig. 17** Radial profiles of the mean streamwise velocity at *x/D* = 5, 10, 15, 20 and 30 with fifth-order polynomial curve fits

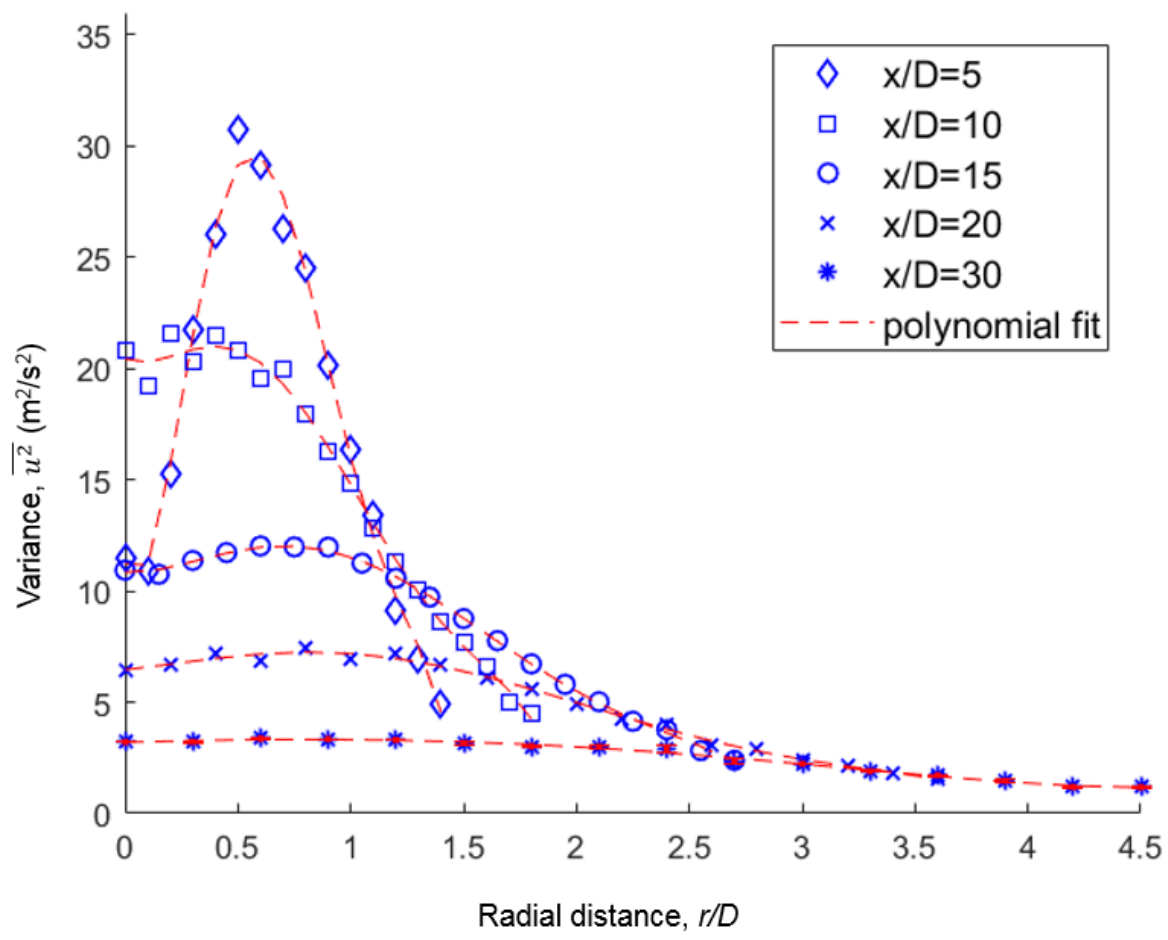

**Fig. 18** Radial profiles of the local streamwise velocity variance at *x/D* = 5, 10, 15, 20 and 30 with fifth-order polynomial curve fits

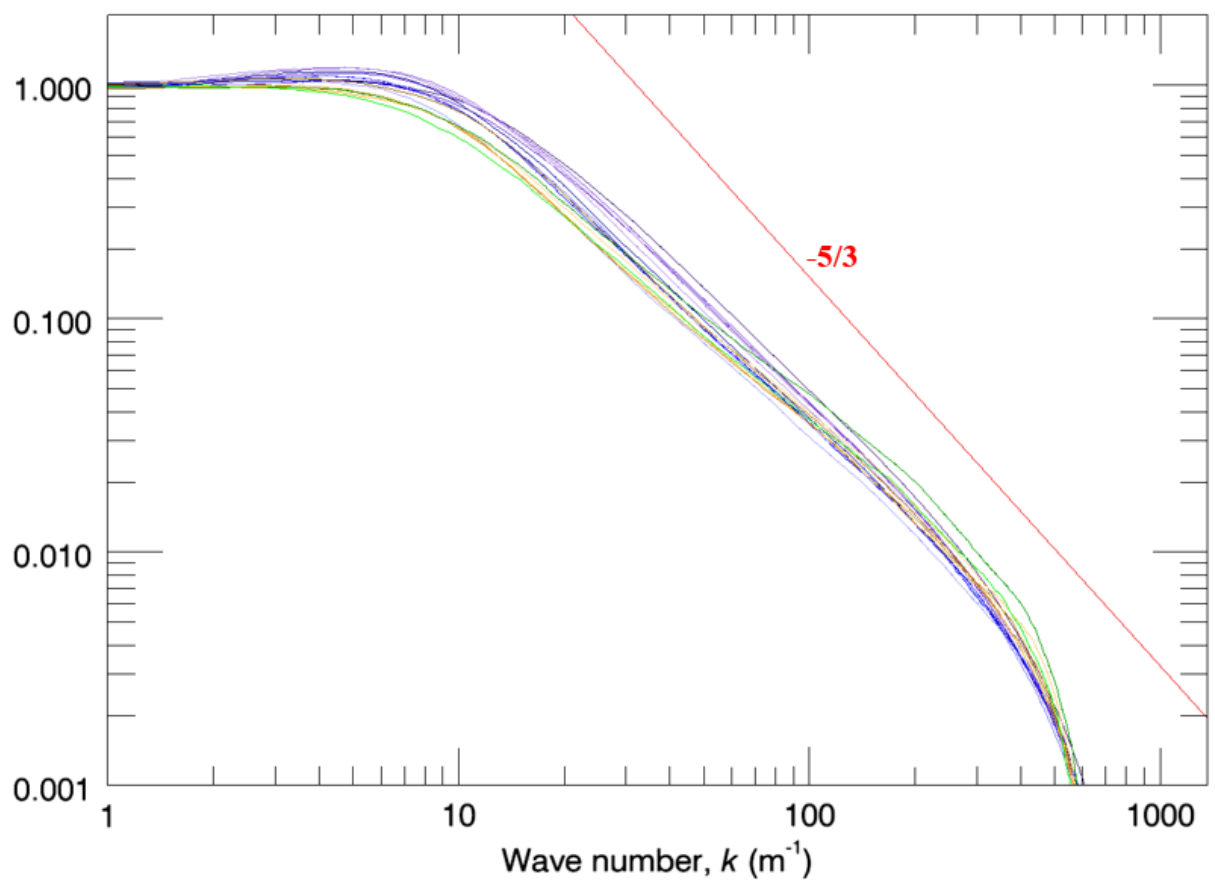

**Fig. 19** Radial development of spatial turbulent kinetic energy spectra (based on the streamwise velocity component) at *x/D* = 30. From *heavy to light purple*: off-axis position 0, 3, 6, 9, 12 mm. From *heavy to light blue*: 15, 18, 21, 24, 27 mm. From *heavy to light brown*: 30, 33, 36, 39 mm. From *heavy to light green*: 42, 45 mm

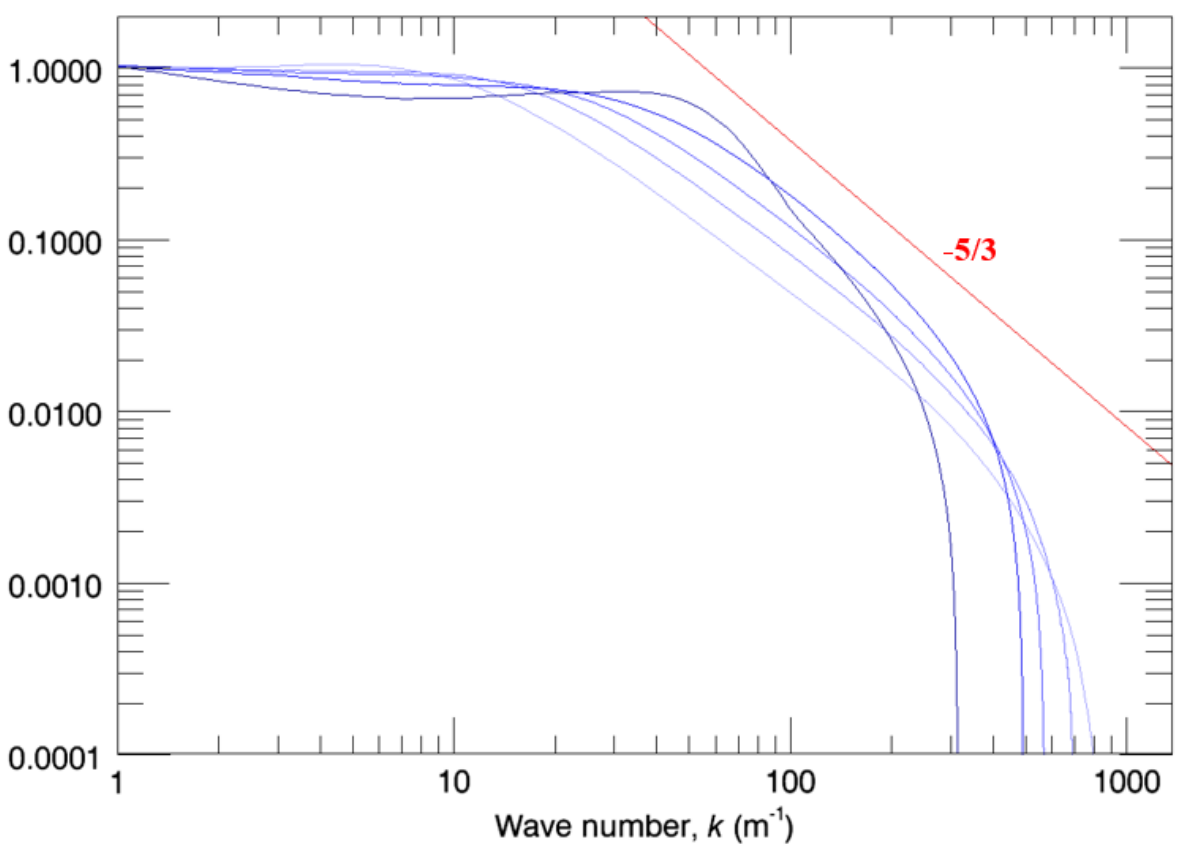

**Fig. 20** Downstream development of spatial turbulent kinetic energy spectra (based on the streamwise velocity component) along the centerline ($r$ = 0). *From heavy to light blue*: *x/D* = 5, 10, 15, 20, 30

## 4 Conclusions

A high dynamic range, transparent and accurate novel LDA system has successfully been developed. Measurement results have been compared to a typical classical hardware driven processor. Detailed diagnostics of the measurements shows that the novel software driven processor can successfully counter several of the limitations of the hardware driven counterpart, including quantization and clipping of the data. Both hardware and software have also been tested and validated for data acquisition and processing, respectively, which provides results that are in a good agreement with previous studies on turbulent round jets, even for challenging large off-axis distances. This proven functionality opens up an opportunity to further investigate the dynamics of significantly more challenging high intensity and high shear flows (in particular potentially non-equilibrium flows) to properly test the well-known local equilibrium hypothesis for the structure of the small scale turbulence.

**Acknowledgements** The authors wish to acknowledge the support of Ministry of Education Malaysia, DTU Mechanical Engineering, Reinholdt W. Jorck og Hustrus Fond (grant journal no. 13-J9-0026), Fabriksejer, Civilingeniør Louis Dreyer Myhrwold og hustru Janne Myhrwolds Fond (grant journal no. 13-M7-0039, 15-M7-0031 and 17-M7-0035) and Siemens A/S Fond grant no. 41. Credits should be also given to Pierre Margotteau, who helped to develop the first generation of the processor. CMV also acknowledges financial support from the European Research council: This project has received funding from the European Research Council (ERC) under the European Unions Horizon 2020 research and innovation program (grant agreement No 803419).

## Appendix I

### List of measurement points for Fig. 6

| | Radial distance [mm] | | | | |
|---|---|---|---|---|---|
| **MP** | ***x/D*=5** | ***x/D*=10** | ***x/D*=15** | ***x/D*=20** | ***x/D*=30** |
| $r_0$ | 0 | 0 | 0 | 0 | 0 |
| $r_1$ | 1 | 1 | 1.5 | 2 | 3 |
| $r_2$ | 2 | 2 | 3 | 4 | 6 |
| $r_3$ | 3 | 3 | 4.5 | 6 | 9 |
| $r_4$ | 4 | 4 | 6 | 8 | 12 |
| $r_5$ | 5 | 5 | 7.5 | 10 | 15 |
| $r_6$ | 6 | 6 | 9 | 12 | 18 |
| $r_7$ | 7 | 7 | 10.5 | 14 | 21 |
| $r_8$ | 8 | 8 | 12 | 16 | 24 |
| $r_9$ | 9 | 9 | 13.5 | 18 | 27 |
| $r_{10}$ | 10 | 10 | 15 | 20 | 30 |
| $r_{11}$ | 11 | 11 | 16.5 | 22 | 33 |
| $r_{12}$ | 12 | 12 | 18 | 24 | 36 |
| $r_{13}$ | 13 | 13 | 19.5 | 26 | 39 |
| $r_{14}$ | 14 | 14 | 21 | 28 | 42 |
| $r_{15}$ | N/A | 15 | 22.5 | 30 | 45 |
| $r_{16}$ | N/A | 16 | 24 | 32 | N/A |
| $r_{17}$ | N/A | 17 | 25.5 | 34 | N/A |
| $r_{18}$ | N/A | 18 | 27 | 36 | N/A |

## Appendix II

### Detailed steps to estimate the momentum integral, *M* at *x/D*=30

The momentum flux per unit mass at the jet exit is calculated by

$$\boldsymbol{M_0 = \pi\left(\frac{D}{2}\right)^2 \bar{u}^2 = 0.0962\ m^4/s^2}$$

At 30 jet exit diameters downstream of the jet nozzle, the momentum across the jet should be the same since the momentum is conserved for a free turbulent jet. The momentum integral to second order can be expressed as

$$\boldsymbol{M = 2\pi\int_0^{\infty}\left[\bar{u}^2 + \overline{u^2} - \frac{1}{2}\left(\overline{v^2} + \overline{w^2}\right)\right] r\, dr} \quad (3)$$

as shown in (Hussein, Capp, and George 1994).

This integral requires knowledge about the second order moments of the two additional components of velocity, namely *v*-variance, $\overline{v^2}$ and *w*-variance, $\overline{w^2}$. It has previously been established that the static statistical moments in a turbulent round jet obey axisymmetry (George and Hussein 1991). Furthermore, (Hussein, Capp, and George 1994) provides usable quantitative data on the relation between the variances of the velocity components. The Reynolds stress profiles of Figure 9-11 in (Hussein, Capp, and George 1994) are digitized and replotted all together in Fig. 21.

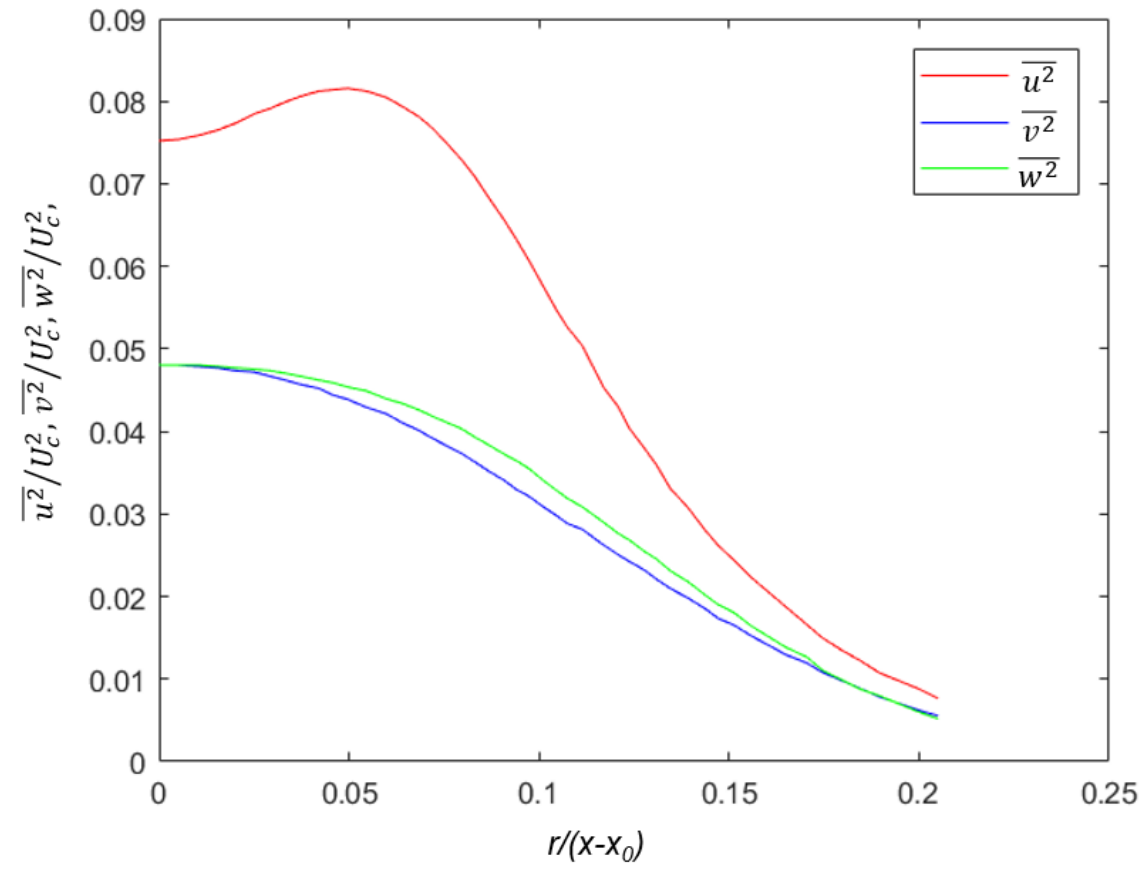

**Fig. 21** Streamwise, radial and azimuthal components of turbulence kinetic energy (normalized by the square of the centreline velocity) at *x/D*=30

The ratio between each pair of profiles has been plotted in Fig. 22. The average ratio between each pair of variance profiles has also been computed accordingly:

$$\boldsymbol{\overline{v^2}/\overline{w^2}} \approx 0.95 \quad (4)$$

$$\boldsymbol{\overline{v^2}/\overline{u^2}} \approx 0.60 \quad (5)$$

$$\boldsymbol{\overline{w^2}/\overline{u^2}} \approx 0.63 \quad (6)$$

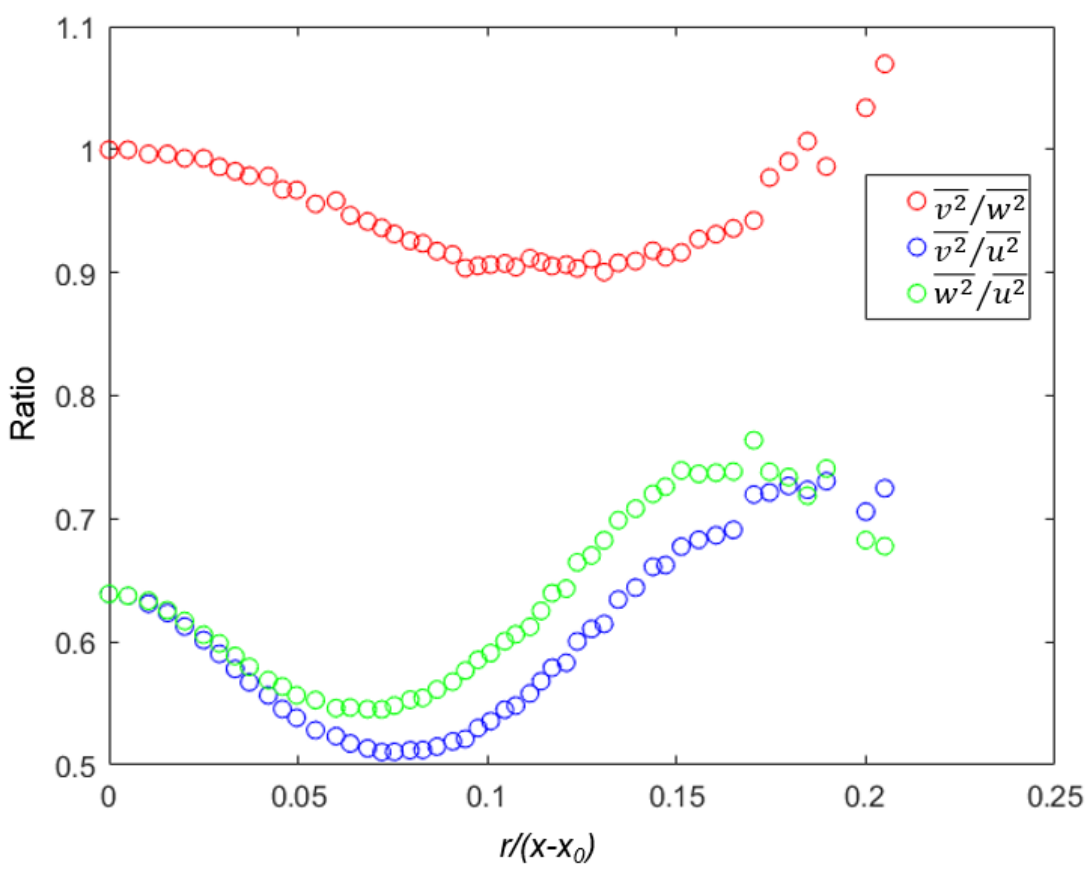

**Fig. 22** Ratio between two different components

Substituting Equation (5) and (6) into Equation (3) resulted in:

$$\boldsymbol{M = 2\pi\left[\int_0^{r_{max}} \bar{u}^2 r\, dr - 0.39\int_0^{r_{max}} \overline{u^2} r\, dr\right]} \quad (7)$$

where $r_{max}$ = 45 mm. The mean streamwise velocity and variance profiles (particularly for *x/D*=30) in the integral are based on the fifth-order polynomial obtained by replotting the profiles over the non-normalized radial distance, *r*, which is given by

$$\overline{\boldsymbol{u}}(\boldsymbol{r}) = (\mathbf{8.53} \times \mathbf{10^7})\, \boldsymbol{r}^5 - (\mathbf{1.13} \times \mathbf{10^7})\, \boldsymbol{r}^4 \quad + (\mathbf{5.88} \times \mathbf{10^5})\, \boldsymbol{r}^3 - (\mathbf{1.4} \times \mathbf{10^4})\boldsymbol{r}^2 = -5.365$$

$$\overline{\boldsymbol{u}^2}(\boldsymbol{r}) = \mathbf{2.35}\, \boldsymbol{r}^5 - \mathbf{0.084}\, \boldsymbol{r}^4 - \mathbf{0.001}\boldsymbol{r}^3 - \mathbf{0.0002}\, \boldsymbol{r}^2 = -\mathbf{4.07}\boldsymbol{x}\mathbf{10} - \mathbf{7}$$

## Affiliations

Mohd Rusdy Yaacob[1, ID]. Rasmus Korslund Schlander[2].
Preben Buchhave[3]. Clara Marika Velte[4, ID]

Mohd Rusdy Yaacob
rusdy@utem.edu.my

Rasmus Korslund Schlander
r.schlander19@imperial.ac.uk

Preben Buchhave
buchhavepreben@gmail.com

Clara Marika Velte
cmve@dtu.dk

[1] Faculty of Electrical Engineering, Universiti Teknikal Malaysia Melaka, Hang Tuah Jaya, 76100 Durian Tunggal, Melaka, Malaysia

[2] Department of Aeronautics, Imperial College London, London SW7 2AZ, United Kingdom

[3] Intarsia Optics, Sønderskovvej 3, 3460 Birkerød, Denmark

[4] Department of Mechanical Engineering, Technical University of Denmark, 2800 Kgs. Lyngby, Denmark